\documentclass[11pt,prd,aps,onecolumn,superscriptaddress,floatfix,notitlepage,nofootinbib,longbibliography]{revtex4-1}
\usepackage[latin1]{inputenc}
\usepackage{mathrsfs}
\usepackage{amsmath,dsfont}
\usepackage{amsfonts,natbib}
\usepackage{amssymb}
\usepackage{bm}
\usepackage[pdftex]{graphicx}
\usepackage{comment}

\usepackage[colorlinks=true,allcolors=blue]{hyperref}

\begin{document}
	\title{Squeezed Quasinormal Modes from Nonlinear Gravitational Effects} 
\author{Sreenath K. Manikandan}
\email{skm@tifrh.res.in}
\affiliation{Nordita, Stockholm University and KTH Royal Institute of Technology, Hannes Alfv\'{e}ns v\"{a}g 12, SE-106 91 Stockholm, Sweden}
\affiliation{Tata Institute of Fundamental Research Hyderabad, 36/P, Gopanpally Village, Serilingampally Mandal, Hyderabad, Telangana 500046, India}
\author{Frank Wilczek}
\email{fwilczek@asu.edu}
\affiliation{Department of Physics, Stockholm University, AlbaNova University Center, 106 91 Stockholm, Sweden}
\affiliation{Department of Physics and Origins Project, Arizona State University, Tempe, Arizona 25287, USA}
\affiliation{T. D. Lee Institute, Shanghai 201210, China}
\affiliation{Wilczek Quantum Center, Department of Physics and Astronomy, Shanghai Jiao Tong University, Shanghai 200240, China }
\affiliation{Center for Theoretical Physics, Massachusetts Institute of Technology, Cambridge, Massachusetts 02139, USA}
\affiliation{Nordita, Stockholm University and KTH Royal Institute of Technology, Hannes Alfv\'{e}ns v\"{a}g 12, SE-106 91 Stockholm, Sweden}
\date{\today}
   \begin{abstract}
       We estimate the degree of squeezing possible in gravitational waves due to nonlinear gravitational effects  in the weakly perturbative regime.  Using the predicted amplitude ratios for higher harmonic generation in the ringdown phase of a black hole merger event, we estimate the relevant degree of squeezing produced by a Schwarzschild singularity to be of the order of one percent.
   \end{abstract}

\maketitle
\section{Introduction}

Consideration of the interaction of radiation with quantum detectors requires some assumption, implicit or explicit, about the quantum nature of the radiation.  Classical treatment of the radiation, amounting to the replacement of its field strength by c-numbers, corresponds implicitly to the assumption that it is embedded in the quantum framework as a coherent state (coherent state hypothesis).  Squeezing takes us outside the domain of coherent states, and brings in characteristically quantum-mechanical effects.  Since nonlinearities at the source or in propagation generically induce squeezing, and general relativity is an inherently nonlinear theory, some amount of squeezing is likely to be present in all forms of gravitational radiation.  It is a challenging task, however, to make quantitative estimates for realistic, complex sources. Here we address this challenge by developing a simple theoretical framework that takes the amplitude ratio of higher order gravitational waves produced by nonlinear gravitational self-interactions in the background field as the input. While our model is generically applicable, we make quantitative estimates for the degree of squeezing using the amplitude ratios for radiation emitted during the ``ringdown'' following black hole formation, which gives a conservative estimate for the relevant degree of nonlinearity.   

Black hole quasi-normal modes~\cite{ReggeWheeler,Zerilli,Vishu,berti_quasinormal_2009,kokkotas_quasi-normal_1999,berti_black_2025} describe the ringdown stage of radiation.  Although their amplitude is relatively weak, they are (relatively) tractable mathematically.  
Indeed, earlier works have estimated the nonlinear-gravitational effects systematically in the observable ringdown signal~\cite{tomita_non-linear_1974,tomita_nonlinear_1976,PhysRevLett.130.081401,gleiser_second-order_1996,gleiser_gravitational_2000,PhysRevLett.130.081402,PhysRevD.76.084007,PhysRevD.76.061503}.  Expansion of Einstein's field equations to the second order about a black hole spacetime, 
\begin{eqnarray}
    \tilde{g}_{\mu\nu}&\approx&
g_{\mu\nu} +h_{\mu\nu}^{(1)}+h_{\mu\nu}^{(2)},\nonumber\\
G_{\mu\nu}(\tilde{g})&\approx & G_{\mu\nu}^{(1)}[h^{(1)}]+G_{\mu\nu}^{(1)}[h^{(2)}]+G_{\mu\nu}^{(2)}[h^{(1)},h^{(1)}]+...
\end{eqnarray}
suggests the following form for dynamical equations that describe the metric perturbations,
\begin{eqnarray}
    G_{\mu\nu}^{(1)}[h^{(1)}] =0,~~\text{and}~~G_{\mu\nu}^{(1)}[h^{(2)}]=-G_{\mu\nu}^{(2)}[h^{(1)},h^{(1)}].\label{eqG}
\end{eqnarray}
Above $G_{\mu\nu}^{(i)},~i=1,2$ refers to the Einstein tensor $G_{\mu\nu}$ expanded to appropriate order~\cite{tomita_non-linear_1974,tomita_nonlinear_1976,PhysRevLett.130.081401,gleiser_second-order_1996,gleiser_gravitational_2000,PhysRevLett.130.081402,PhysRevD.76.084007,PhysRevD.76.061503}.  In the Schwarzschild case, the first of the above equations leads to Regge-Wheeler~\cite{ReggeWheeler}, and Zerilli~\cite{Zerilli} equations for odd and even parity black hole quasi-normal modes, while the second equation leads to similar equations with a source term that is, up to derivatives, quadratic in the first order metric perturbations~\cite{PhysRevLett.130.081401,PhysRevLett.130.081402,PhysRevD.76.084007,PhysRevD.76.061503}. 

The suggestion to use the nonlinearities in black hole ringdown as a probe for quantum gravity effects, based on the analogy to second harmonic generation in quantum optics, was also made in Ref.~\cite{guerreiro_nonlinearities_2023}. Following a systematic analysis using simple dimensional arguments, we show that the degree of squeezing for such quadratic gravitational self-interactions turns out to be the square of the amplitude ratios. Using these, we predict an order of one percent squeezing of the phase-space quadratures in this tractable perturbative regime.   
In interesting limits we also relate our problem to a parametrically excited oscillator, a context in which squeezing has been quantified both theoretically and experimentally~\cite{chaturvedi_limits_2002,mallet_quantum_2011,murch_reduction_2013,wu_squeezed_1987,adamyan_strong_2015,SofiaSree,das2025squeezed}. Here again, our analysis shows that the estimated degree of squeezing is of the order of the square of the amplitude ratio between the nonlinearly produced component and the fundamental mode.  Our work complements  recent discussions of how the coherent state hypothesis for gravitational radiation might be tested in practice with LIGO-type interferometers~\cite{ParikhEPJD,ParikhPRL,ParikhPRD}, and especially resonant mass detectors~\cite{tobar_detecting_2024,manikandan_complementary_2025,manikandan_probing_2025,tobar_detecting_2024,Manikandan_Wilczek_acoherence,shenderov_stimulated_2024}. 

\section{Quantum mechanical description of nonlinear gravitational wave interactions}
Gravitational radiation, owing to inherent nonlinearities of gravity, is expected to have some degree of squeezing. Estimating the degree of squeezing involved is the challenging task. However gravitational wave scenarios where the relevant degree of nonlinearity can be estimated systematically gives hope that the degree of squeezing involved is both non-trivial and predictable. Higher harmonic generation in quasinormal modes provide one such framework which gives a conservative estimate for the relevant degree of nonlinearity involved in terms of calculable amplitude ratios. Several recent works have noted that the coupled nature of wave equations at the second order imply higher harmonic generation of black hole driven quasinormal modes, with leading contributions to the $l=4,m=4$ second-order quasinormal modes sourced by the $l=2,m=2$ mode~\cite{tomita_non-linear_1974,tomita_nonlinear_1976,PhysRevLett.130.081401,gleiser_second-order_1996,gleiser_gravitational_2000,PhysRevLett.130.081402,PhysRevD.76.084007,PhysRevD.76.061503}.
 The ``quasinormal" aspect reflects the fact that the mode frequencies involved are complex.  
 
In the forthcoming discussions, we consider a toy model inspired by these developments, where we treat the frequencies involved approximately real. Strictly speaking, this approximation is only valid to describe early-time behavior of long-lived quasinormal modes. Therefore much of what follows can be understood as a rigorous toy model for normal mode gravitational waves with nonlinear self-interactions. Nevertheless, we keep the mode labels from the quasinormal mode context in all our forthcoming discussions, since we eventually would like to estimate the degree of squeezing in terms of the relevant nonlinearities in a gravitational scenario, which we extract from the amplitude ratios predicted for higher harmonic generation in quasinormal modes. There are good prospects to observe second harmonic generation sourced by lower order quasinormal modes, since the expected conversion ratio for lowest-order modes of non-rotating black holes is approximately, $\frac{|h_{44q}|}{|h_{220}^2|}\sim 0.15$~\cite{PhysRevLett.130.081401,PhysRevLett.130.081402,PhysRevD.76.084007,PhysRevD.76.061503}. Perturbation theory for rotating (Kerr) spacetimes, notably developed by Teukolsky~\cite{Teukolsky} is more involved, but here too higher-harmonic quasinormal modes arise~\cite{quadraticQNMKerr}. Since quasinormal modes are decaying modes, we expect that the  amplitude ratios predicted here provide a conservative estimate for the relevant degree of nonlinearity in comparable normal mode settings as well.  

 To the first order, the  $l=2,m=2$ fundamental mode obeys the following wave equation,
\begin{equation}
\ddot{h}_{220}+\omega_{220}^2h_{220}= 0\label{eqFunda}.
\end{equation}
Up to second order, a temporal wave equation describing the second harmonic generation of modes will have the generic form
\begin{equation}
\ddot{h}_{44q}+\omega_{44j}^2h_{44q}\approx\sqrt{2}\Lambda_j h_{220}^2\label{eqSeco},
\end{equation}
where $\Lambda_j$ with $[\Lambda_j] = \left[T^{-2}\right]$ is an effective coupling to be determined. A framework to derive this coupled wave equation in the quasinormal mode context is presented in Appendix~\ref{QNM} where some of the challenges facing a rigorous derivation are also outlined.  The frequency $\omega_{44j}$ appearing in the above equation exemplifies that, in the context of quasinormal modes, the modes that get pushed to second harmonic are the $l=4,m=4$ modes fundamental modes, sourced by the $l=2,m=2$ mode to the leading order. When the self interaction $\Lambda_j$ is neglected, the equation correspond to a first order quasinormal mode with $l=4,m=4$ that exist in the original spectrum.
Looking at the form of the wave equation, and using Fourier analysis, one can conclude that the forced second harmonic mode will actually oscillate at twice the fundamental mode frequency. Alternatively, assuming a solution $h_{220}(t)\approx a\text{Re}(e^{-i\omega_{220}t})=  a \cos{(\omega_{220,R}t)}e^{-\omega_{220,I}t}\approx   a \cos{(\omega_{220}t)}$, for the fundamental mode [where we have also neglected damping, $\omega_{220}\approx \text{Re}(\omega_{220})$], we obtain that the particular solution to the above equation has the following form,
\begin{equation}
    h_{44j}^{(p)}(t)=\frac{a^2\Lambda_j}{\sqrt{2}\omega_{44j}^2}+\frac{a^2 \Lambda_j \cos(2\omega_{220}t)}{\sqrt{2}(\omega_{44j}^2-4\omega_{220}^2)}.\label{solPart}
\end{equation}
Note that the temporal part of the particular solution oscillates at frequency $\omega_{44}^{(p)}=2\omega_{220}$, which is the second harmonic generated mode due to the effective forcing by the source term.   

To compute quantum effects in a meaningful way it is important to estimate the effective interaction Hamiltonian in the quantum description, considering gravitational wave frequencies that are approximately real. As mentioned, long-lived quasinormal modes can very well be approximated in this manner. The quadratically sourced wave equation, Eq.~\eqref{eqSeco} can be derived from  the lagrangean,
\begin{eqnarray}
    L_{q} = \left[\frac{1}{2}\left(\dot{h}_{44q}\right)^2-\frac{\omega_{44j}^2}{2}\left(h_{44q}\right)^2+\sqrt{2}\Lambda_j h_{220}^2h_{44q}\right].\label{Lagra}
\end{eqnarray}
To this, we should add the lagrangean for the fundamental mode for completeness, given by,
\begin{equation}
    L_{0} =  \frac{1}{2}\dot{h}_{220}^2-\frac{\omega_{220}^2}{2}h_{220}^2,
\end{equation}
which suggests that some sort of back-reaction to the fundamental mode is inevitable in the presence of nonlinear interactions. It also helps to note that all the $44j$ modes could get pushed to the second harmonic to different degrees, determined by the coupling strengths $\Lambda_j$, but we expect that the near-resonant lower-$j$ modes would be the ones that will get maximally pushed. We can therefore replace $\Lambda_j$ with an effective coupling $\Lambda$ in the forthcoming discussions.  

The wave equation for the fundamental mode also modifies to account for the non-linear interactions. Variation of the total Lagrangian $L=L_{q}+L_{0}$ with respect to $\dot{h}_{220}$ yields,
\begin{equation}
\ddot{h}_{220}+\left[\omega_{220}^2-2\sqrt{2}\Lambda h_{44q}\right]h_{220} = 0.
\end{equation}
We see that the mode $h_{44q}$ modulates the frequency of the fundamental mode in a time-dependent manner. Since the temporal part of $h_{44q}$ oscillates at frequency $\omega_{p}=2\omega_{220}$, we can approximate it as $h_{44q}(t)\approx h_{0}^{(q)} \cos(2\omega_{220}t)$. With this, we note that the equation governing the fundamental mode is modified as,
\begin{equation}
\ddot{h}_{220}+\left[\omega_{220}^2-2\epsilon\cos{(2\omega_{220}t)} \right]h_{220} = 0,
\end{equation}
where $\epsilon = \sqrt{2}\Lambda h_{0}^{(q)}$. This is Mathieu's equation in the standard form that describes a parametrically driven oscillator with $\omega_p=2\omega_{220}$. Considering the corresponding quantum oscillator problem, we note that such a drive naturally squeezes the fundamental mode as it can be understood as time-dependent modulations of the quantum harmonic potential, which has a squeezing effect~\cite{chaturvedi_limits_2002,mallet_quantum_2011,murch_reduction_2013,wu_squeezed_1987,adamyan_strong_2015,SofiaSree,das2025squeezed}. The corresponding degree of squeezing, to the leading order, increases linearly with time~\cite{SofiaSree,das2025squeezed},
\begin{equation}
    r_{\text{sq}}(t) \approx \epsilon t/(2\omega_{220}) = \Lambda h_0^{(p)}t/(\sqrt{2}\omega_{220})\label{paramSq}. 
\end{equation}
An initial coherent state gets squeezed as,
\begin{equation}
    \langle (\Delta X)^2\rangle = \frac{1}{2}e^{-2r} = \frac{1}{2}(1-2r)\approx \frac{1}{2}-r.
\end{equation}
We estimate the degree of squeezing it predicts in Sec.~\ref{Sec:squeezing}.

The overtone mode attaches a primary degree of freedom of the quantized radiation field well outside the black hole to the nonlinear driving term that excites it.  Similar situations have been treated in quantum optics and quantum acoustics, where the feedback from such modes has been shown to induce squeezing in the primary mode.  Informed and inspired by that work, in the remainder of this section we provide a more systematic estimate in the framework of quantum theory, through construction of an effective Hamiltonian. 

\subsection{Effective Hamiltonian and dimensional analysis}
An alternative semi-quantitative approach to estimate the degree of squeezing can be obtained with the help of dimensional analysis. We begin by noticing that both Eq.~\eqref{eqFunda} and Eq.~\eqref{eqSeco} we started with are for the strain amplitudes which are dimensionless. In order to map to quantum harmonic oscillators in the corresponding quantum theory and squeezing at appropriate time scales of interest, it is important to understand what is the effective quadrupole moment of these oscillators in a corresponding quantum description. This can be achieved by promoting the total lagrangean to one that has the dimensions of energy. To this end, we note that the lagrangean must be multiplied by a quantity that has dimensions of a mass quadrupole moment, $[\mu] = [ML^2]$:
\begin{eqnarray}
    \mu L_{q} = \mu\left[\frac{1}{2}\left(\dot{h}_{44q}\right)^2-\frac{\omega_{44j}^2}{2}\left(h_{44q}\right)^2+\sqrt{2}\Lambda h_{220}^2h_{44q}\right],
\end{eqnarray}
such that,
\begin{equation}
    \left[\mu L_{q}\right] = [ML^2T^{-2}],
\end{equation} has the dimensions of energy. Given that gravitational radiation is quadrupolar in nature, and sources that create are time-varying quadrupole moments, we can predict the scaling of $\mu$ with fundamental constants, and the frequency of the mode to be,
\begin{equation}
    [\mu] = \left[\frac{c^5}{G\omega^3}\right].
\end{equation} To estimate $\mu$ more definitively, we note that the corresponding wave equation for the second harmonic mode have the following form,
\begin{equation}
\mu\ddot{h}_{44q}+\mu\omega_{44j}^2h_{44q} \approx \sqrt{2}\Lambda\mu h_{220}^2.
\end{equation}
Here we can fix  $\mu\sim\frac{c^5}{16\pi G\Omega^3}$ such that the kinetic term of the Hamiltonian that yields the inertial term of the above equation maps to the energy of the wave in the mode volume, $V=(c/\Omega)^3$ given by $(E/V)(c/\Omega)^3\sim \frac{c^2}{32\pi G}\langle \dot{h}^2\rangle (c/\Omega)^3$, averaged over cycles.   

Now the corresponding quantum harmonic oscillator Hamiltonian can be quantized in the usual manner, which correspond to promoting the field quadratures to operators in the following standard procedure,
\begin{equation}
    h_{\omega_j}(t)\rightarrow \hat{h}_{\omega_j}(t) = \sqrt{\frac{\hbar}{2\mu_j\omega_j}}(A_{\omega_j} e^{-i\omega_j t}+A_{\omega_j}^\dagger e^{i\omega_j t}) =\sqrt{\frac{8\pi G\hbar\omega_j^2}{c^5}}(A_{\omega_j} e^{-i\omega_j t}+A_{\omega_j}^\dagger e^{i\omega_j t}) 
\end{equation}

From the lagrangean, we can also estimate the corresponding interaction Hamiltonian. Making use of the simplified notation $h_{220}\equiv h_{\omega}$ and $h_{44q}\equiv h_{2\omega}$, we see that the interaction hamiltonian will have the form,
\begin{eqnarray}
    H_I &\approx& -\sqrt{2}\Lambda\mu\left[\hat{h}_{\omega}(t)\right]^2\hat{h}_{2\omega}(t) \nonumber\\&=& -\hbar\sqrt{\hbar}\Lambda\mu\omega^3 \left(\frac{16\pi G}{c^5}\right)^{3/2} (A_{\omega}e^{-i\omega t}+A_{\omega}^\dagger e^{i\omega t})^2(A_{2\omega}e^{-i2\omega t}+A_{2\omega}^\dagger e^{i2\omega t})\nonumber\\
    &=&-\frac{\hbar\lambda}{2} (A_{\omega}e^{-i\omega t}+A_{\omega}^\dagger e^{i\omega t})^2(A_{2\omega}e^{-i2\omega t}+A_{2\omega}^\dagger e^{i2\omega t}),
\end{eqnarray} Since the mapping concerns the kinetic (or inertial, term), it also makes sense to evaluate $\mu$ at frequency $\Omega = \omega_{44q}$ for the coupling, since the particular solution $h_{44q}(t)$ is a (damped) sinusoidal wave at frequency $\omega_{44q}=2\omega_{220}$.
Substituting the approximate value for $\mu$ evaluated at the frequency of the second Harmonic mode,
 $\mu\approx\frac{c^5}{16\pi G(2\omega)^3}$, we obtain, 
\begin{equation}
    \lambda = \Lambda\sqrt{\frac{\pi G\hbar }{c^5}}=\Lambda \tau_p\sqrt{\pi}.
\end{equation}
where $\tau_p=\sqrt{\frac{G\hbar }{c^5}}$ is the Planck time.

To relate to some of the standard discussions of second harmonic generation in quantum optics, we can make a further canonical transformation,
\begin{equation}
    \frac{1}{\sqrt{2}}\left(A_{\omega_p}e^{i\omega_p t}+A_{\omega_p}^\dagger e^{-i\omega_p t}\right)\rightarrow \frac{i}{\sqrt{2}}\left(A_{\omega_p}e^{i\omega_p t}-A_{\omega_p}^\dagger e^{-i\omega_p t}\right) 
\end{equation} and \begin{equation}
    \frac{1}{\sqrt{2}i}\left(A_{\omega_p}e^{i\omega_p t}-A_{\omega_p}^\dagger e^{-i\omega_p t}\right)\rightarrow \frac{1}{\sqrt{2}}\left(A_{\omega_p}e^{i\omega_p t}+A_{\omega_p}^\dagger e^{-i\omega_p t}\right), 
\end{equation} so that the interaction Hamiltonian can be represented as,
\begin{equation}
    H_I \approx -i\frac{\hbar\lambda}{2}\left[(A_{\omega}^\dagger)^2e^{-2i\omega_0 t}+A_{\omega}^2e^{2i\omega_0 t}+2A_{\omega}^\dagger A_{\omega}+1\right]\left(A_{\omega_p}e^{i\omega_p t}-A_{\omega_p}^\dagger e^{-i\omega_p t}\right).
\end{equation}
Noting that $\omega_p=2\omega_0$ (denoted as $2\omega$ in subscripts), we can make the rotating wave approximation to neglect the terms that oscillate at $\pm 2\omega_0$ and $\pm 4\omega_0$, and only keep terms that are energy conserving. Finally we obtain,
\begin{equation}
    H_I \approx -i\frac{\hbar\lambda}{2}\left[(A_{\omega}^\dagger)^2A_{2\omega}-A_{\omega}^2A_{2\omega}^\dagger\right].\label{HamSHG}
\end{equation}

\subsection{Estimating the coupling $\lambda$ from phenomenology}
Above we have reduced the problem to a single effective parameter $\lambda$. Here we describe a phenomenological model that allows for estimating the coupling $\lambda$ from observable amplitudes in rather generic scenarios. To this end, we note that the amplitude $\frac{\lambda}{2}$ is determined by the fraction of energy dissipated via second harmonic generation, which numerical general relativity provides good estimates, that the amplitude of the second harmonic mode can be $\sim 10\%$ to that of the fundamental mode in some scenarios~\cite{PhysRevLett.130.081401,PhysRevLett.130.081402}. In fact, we can take the semi-classical limit to treat the fundamental mode in a coherent state $A_{\omega}|\alpha_\omega\rangle = \alpha_\omega|\alpha_\omega\rangle$, and approximate the interaction Hamiltonian as,
\begin{equation}
     H_I \approx -i\frac{\hbar\lambda}{2}\left[(\alpha_\omega^*)^2A_{2\omega}-\alpha_\omega^2A_{2\omega}^\dagger\right].
\end{equation}
With these approximations, also neglecting the depletion of the fundamental mode, the effective dynamics reduces to a displacement operator on the second harmonic mode, which prepares a coherent state of the second harmonic mode with the amplitude,
\begin{equation}
    |\alpha_{2\omega}| \approx \left|\frac{\lambda}{2}\right||\alpha_\omega|^2 t.\label{amp2}
\end{equation}
Note that the above relation holds for the quantum mechanical amplitude of the second harmonic mode, not to be confused with its temporal mode profile. The temporal mode profile for the second harmonic mode oscillates at frequency $2\omega$ as indicated by the subscript. As expected from Eq.~\eqref{solPart}, the amplitude is proportional to the coupling $\Lambda$ as well as the square of the amplitude of the fundamental mode. One can also identify a timescale $t\sim 1/\omega$ that can be chosen such that the amplitudes agree.

From Eq.~\eqref{amp2}, it seems appropriate that we could determine $\lambda$ at short times from simulations or actual data as,
\begin{equation}
    |\lambda|= \lim_{\Delta t\rightarrow 0}\frac{2}{\Delta t}\frac{|\alpha_{2\omega}(\Delta t)|-|\alpha_{2\omega}(0)|}{|\alpha_\omega|^2}.
\end{equation}
Numerical relativity can predict the classical gravitational wave amplitude ratios at early times, and assuming gravitational radiation quantized within a cubical box of reduced volume $V=(c/\omega)^3$, the classical gravitational wave amplitudes can be related to the coherent state amplitudes at early times as (see Appendix.~\ref{AppA}), 
\begin{eqnarray}
      \frac{h_{2\omega}}{h_{\omega}^2} \approx \sqrt{\frac{c^5}{8\pi G\hbar\omega^2}}\frac{|\alpha_{2\omega}|}{|\alpha_{\omega}|^2} \approx \sqrt{\frac{c^5}{8\pi G\hbar\omega^2}}\left|\frac{\lambda}{2}\right|t.\label{ampRatio}
\end{eqnarray} 
We can use this amplitude ratio to estimate the degree of squeezing in the fundamental mode.
\section{Degree of Squeezing\label{Sec:squeezing}}
The interaction Hamiltonian in Eq.~\eqref{HamSHG} describes second harmonic generation, which is a well-known method in quantum optics to generate and probe sub-Poissonian states of the radiation field~\cite{MANDEL1982437}. We can extend the analysis to the gravitational context as well, and proceed to estimate the degree of squeezing. To this end, we can use the amplitude ratios predicted for the ringdown phase to estimate the relevant nonlinearity in a weakly perturbative regime of quantum gravity. Note that the Hamiltonian from Eq.~\eqref{HamSHG} leads to the following set of equations that describe second harmonic generation,
\begin{equation}
    \frac{dA_{\omega}}{dt} =-\lambda  A^\dagger_\omega A_{2\omega},~~\frac{dA_{2\omega}}{dt} = \frac{\lambda}{2}A_{\omega}^2.
\end{equation}
Now the analysis of squeezing can be carried out systematically, where we take $\lambda$ to be real for simplicity. With the fundamental mode initially in a coherent state $|\alpha_\omega\rangle$ and the second harmonic mode in its vacuum state, to leading order ($\lambda |\alpha_{\omega}| t\ll 1$), the covariance of the fundamental mode changes as, 
\begin{eqnarray}
    \langle (\Delta \hat{X}_{\omega})^2\rangle \approx\frac{1}{2}-\lambda^2t^2|\alpha_\omega|^2/4,~~\text{and}~~\langle (\Delta \hat{P}_{\omega})^2\rangle \approx\frac{1}{2}+\lambda^2t^2|\alpha_\omega|^2/4,\label{pertSqueeze}
\end{eqnarray}
that suggests that squeezing of the sub-Poissonian nature~\cite{MANDEL1982437}. We can also use a perturbatively exact analysis based on Refs.~\cite {OuSHG,PhysRevA.49.2157,suhara_theoretical_1996}, revisited in Appendix.~\ref{AppSHG}. For coherent state initial conditions for the fundamental mode and second harmonic mode in vacuum, where $\langle \hat{X}^2_{j\omega}(0)\rangle=\langle \hat{P}^2_{j\omega}(0)\rangle=\frac{1}{2}$,
one obtains that $
    \langle \hat{X}^2_{j\omega}(\tau)\rangle =  \langle \hat{X}^2_{j\omega}(0)\rangle S_{j\omega X},~~\text{and}~~\langle \hat{P}^2_{j\omega}(\tau)\rangle =  \langle \hat{P}^2_{j\omega}(0)\rangle S_{j\omega P},~~j=1,2.
$
 Here, $\tau = \frac{1}{\sqrt{2}}\lambda |\langle A_{\omega}(0)\rangle|t = \frac{1}{\sqrt{2}}\lambda|\alpha_{\omega}(0)|t$ and,
\begin{eqnarray}
    S_{\omega X} &=&(1-\tau\tanh\tau)^2\text{sech}^2\tau+2\tanh^2\tau \text{sech}^2\tau,\nonumber\\
      S_{2\omega X} &=&\frac{1}{2}(\tanh\tau+\tau\text{sech}^2\tau)^2+\text{sech}^4\tau,\nonumber\\
    S_{\omega P} &=&\frac{1}{2}(\sinh\tau+\tau\text{sech}\tau)^2+\text{sech}^2\tau,\nonumber\\  
     S_{2\omega P}&=&2\tanh^2\tau+(1-\tau\tanh\tau)^2,
\end{eqnarray}
which reduces to Eq.~\eqref{pertSqueeze} at small $\tau$. For $\tau\gg 1$, one obtains that, $S_{\omega X}\rightarrow 4\tau^2 e^{-2\tau}$, $ S_{2\omega X}\rightarrow 1/2$, $S_{\omega P}\rightarrow e^{2\tau}/8$, and $S_{2\omega P}\rightarrow \tau^2$~\cite {OuSHG,PhysRevA.49.2157,suhara_theoretical_1996}.  This suggests substantial squeezing can in principle be generated through this mechanism, especially an exponential enhancement in the squeezing of the $\hat{P}$ quadrature of the fundamental mode. 

To make some estimates relevant to our scenarios, note that we can approximate,
\begin{equation}
    |\tau|= \frac{1}{\sqrt{2}}\lambda|\alpha_{\omega}(0)|t \approx \sqrt{\frac{16\pi G\hbar\omega^2}{c^5}}\left|\frac{h_{2\omega}}{h_{\omega}^2}\right||\alpha_{\omega}(0)|,
\end{equation}
 where we have used Eq.~\eqref{ampRatio}. For $h_\omega$ being the amplitude of the fundamental mode at the source, $|\alpha_\omega(0)|\approx h_\omega\sqrt{\frac{c^5}{32\pi G\hbar\omega^2}}$. Hence we obtain,
\begin{equation}
    |\tau| = \sqrt{\frac{16\pi G\hbar\omega^2}{c^5}}\left|\frac{h_{2\omega}}{h_{\omega}^2}\right|\sqrt{\frac{c^5}{32\pi G\hbar\omega^2}}h_\omega\approx \frac{1}{\sqrt{2}}\left|\frac{h_{2\omega}}{h_{\omega}^2}\right|h_{\omega}\approx  0.1 h_\omega,
\end{equation}
since $\left|\frac{h_{2\omega}}{h_{\omega}^2}\right|\sim 0.15$. While the amplitude of the gravitational radiation is substantially small at our detectors, $h_{\omega}^{D}\approx 10^{-22}$, at the source it can be substantially high, and can be of the order of unity, suggesting that $|\tau|\approx 0.1$. Assuming that the fundamental mode is in a coherent state, we are also in the regime of validity of Eq.~\eqref{pertSqueeze},
\begin{eqnarray}
    \langle (\Delta \hat{X}_{\omega})^2\rangle &\approx&\frac{1}{2}-\lambda^2t^2|\alpha_\omega|^2/4=\frac{1}{2}\left( 1-|\tau|^2\right)=0.495,~~\text{and}\nonumber\\\langle (\Delta \hat{P}_{\omega})^2\rangle &\approx&\frac{1}{2}+\lambda^2t^2|\alpha_\omega|^2/4=\frac{1}{2}\left( 1+|\tau|^2\right)=0.505.\label{Sqest}
\end{eqnarray}
We thus estimate $S_{\omega X}\approx 0.99,~S_{\omega P}\approx 1.01$. This corresponds to a $1\%$ deviation in the noise of the fundamental mode due to second harmonic generation. While small, it is not minuscule.  It serves as a sort of existence proof, showing the existence of a significant quantum effect in the tractable perturbative regime. 

Based on the mapping to the parametrically excited oscillator, we can also make some upper-bound for the degree of squeezing using Eq.~\eqref{paramSq}, assuming the time $t$ is approximately the time it takes for the second harmonic amplitude to go from $0$ to $h_{2\omega}$, and keeping the amplitude $h_{0}^{(p)}$ to be the maximum achievable amplitude in this time. We find that,
\begin{equation}
    \lambda t =\Lambda t \sqrt{\frac{\pi G\hbar}{c^5}}\implies \Lambda t =\sqrt{\frac{c^5 }{\pi G\hbar}}\lambda t = 2\frac{h_{2\omega}}{h_\omega^2}\sqrt{\frac{8\pi G\hbar \omega^2}{c^5}}\sqrt{\frac{c^5 }{\pi G\hbar}}=4\sqrt{2}\omega \frac{h_{2\omega}}{h_\omega^2}
\end{equation}
This yields $  r = \Lambda h_0^{(p)}t/(\sqrt{2}\omega) =4 \frac{h_{2\omega}^2}{h_\omega^2}$.
The actual achievable squeezing will be smaller than this, given that the drive amplitude $h_0^{(p)}$ gradually increases from  $0$ to $h_{2\omega}$. The more careful estimate from Eq.~\eqref{Sqest} yields,
\begin{equation}
    r=|\tau|^2/2 =\frac{1}{4}\frac{h_{2\omega}^2}{h_\omega^2}.
\end{equation}
We note that up to a numerical pre-factor the degree of squeezing estimated agrees with our estimate based on mapping to a parametric oscillator.  Probably the parametric oscillator  analysis overestimates the degree of squeezing: it assumes that that the second harmonic mode of predicted amplitude is available for the whole duration, while in practice it is only being generated due to non-linear interactions.

For reasons of concreteness and simplicity, we focused on a scenario where the fundamental mode begins in a coherent state.   The merger dynamics may well prepare them in an acoherent quantum state, however. We defer a detailed analysis of this to future work. 

\section{Discussion}

We have estimated the degree of squeezing generated in gravitational waves from non-linear gravitational self-interactions. The estimates are made using predicted amplitude ratios from the perturbative regime following the merger event. For this scenario, where only weak nonlinearities are involved,  we find that the predicted higher harmonic generation of quasinormal modes suggests a $\sim 1\%$ squeezing. As mentioned immediately above, our simplifying assumptions about initial conditions are not unlikely to be overly pessimistic. More generally, earlier stages of radiation close to merger, where nonlinearities are stronger, can be expected to yield larger effects.  Recent work suggests possibilities for testing these ideas experimentally~\cite {manikandan_complementary_2025,manikandan_probing_2025,Manikandan_Wilczek_acoherence,ParikhEPJD,ParikhPRD,ParikhPRL,tobar_detecting_2024,shenderov_stimulated_2024}.

Several other astrophysical scenarios have also been suggested as potential candidates to generate quantum mechanical squeezed states of gravitational radiation. They include head-on collisions of black holes~\cite{lovas_quantization_2001} as well as the expansion of the universe, which could squeeze primordial gravitational waves~\cite{grishchuk_quantum_1989}.

We conclude that observing characteristically quantum behavior in gravitational radiation is a challenging but potentially achievable goal. 

\textit{Note added---.} 
A paper recently posted to the arXiv reaches similar conclusions to ours in a different, but closely related problem involving gravitational radiation~\cite{kanno_quantum_2025}.
\section{Acknowledgments}
FW is supported by the U.S. Department of Energy under grant Contract Number DE-SC0012567 and by the Swedish Research Council under Contract No. 335-2014-7424. SKM was supported in part by the Swedish Research Council under Contract No. 335-2014-7424, and the Wallenberg Initiative on Networks and Quantum Information (WINQ). SKM also acknowledges the support from the Department of Atomic Energy, Government of India, under Project Identification No. RTI4007.  We thank Maulik Parikh and Igor Pikovski for stimulating discussions around these subjects.  
\appendix
\section{Classical and quantum amplitudes of gravitational radiation\label{AppA}}
Quantized gravitational radiation within a cubical box of reduced volume $V=(c/\omega)^3$ in the linearized theory could be represented as (assuming a single polarization),
\begin{eqnarray}
    h(t) &=&\frac{1}{\sqrt{2}}\sum_\omega \sqrt{\frac{16\pi G\hbar}{V c^2\omega}}(A_{\omega}e^{-i\omega t}+A_{\omega}^\dagger e^{i\omega t}) = \frac{1}{\sqrt{2}}\sum_\omega \sqrt{\frac{16\pi G\hbar\omega^2}{c^5}}(A_{\omega}e^{-i\omega t}+A_{\omega}^\dagger e^{i\omega t}).
\end{eqnarray}
Assuming that the field is in a coherent state $|\alpha\rangle$, we can relate the number of gravitons to the intensity of the field (accounting for both polarizations),
\begin{eqnarray}
 h_0^2\approx \frac{32\pi G\hbar \omega^2}{c^5}|\alpha_{\omega}|^2.
\end{eqnarray}
Or alternatively, the number of gravitons,
\begin{equation}
    N\approx |\alpha|^2\approx \frac{h_{0}^2c^5}{32\pi G\hbar \omega^2}.
\end{equation}
Based on this, we can estimate the amplitude ratio,
\begin{eqnarray}
    \frac{h_{2\omega}}{h_{\omega}^2} \approx \sqrt{\frac{c^5}{8\pi G\hbar\omega^2}}\frac{|\alpha_{2\omega}|}{|\alpha_{\omega}|^2}.
\end{eqnarray}
\section{An effective approach to obtain the coupled wave equation in the temporal domain for black-hole quasinormal modes\label{QNM}}
Consider the following wave equation that naturally arise in the context of second-order quasinormal modes~\cite{PhysRevLett.130.081401,PhysRevLett.130.081402,PhysRevD.76.084007,PhysRevD.76.061503}:
\begin{equation}
    \left[
\partial_t^2-\partial_x^2+\hat{V}\right]\psi = S.\label{eqRGS}
\end{equation}
The homogeneous part of this equation corresponds to the homogeneous part of Eqns.~\eqref{eqG}, after separating out the angular variables using tensor harmonics. The potential $\hat{V}$ depends on whether we are considering scalar, electromagnetic (vector) or gravitational (tensor) perturbations. Our interest is in the gravitational, spin-two perturbations. In the non-rotating (Schwarzschild) context, the form of the equation or the potential can be broadly classified further in terms of the parity of the solutions. For the even parity (Zerilli~\cite{Zerilli}), solution change by $(-1)^{l}$ under the transformation $(\theta,\phi)\rightarrow (\theta -\pi,\phi+\pi)$, and for the odd parity perturbation (Regge-Wheeler~\cite{ReggeWheeler}), the solution change by $(-1)^{l+1}$ under the transformation $(\theta,\phi)\rightarrow (\theta -\pi,\phi+\pi)$. Since the potential is also dependent on $l$, we have kept $\hat{V}$ as an operator such that $\left[\partial_t^2-\partial_x^2+\hat{V}\right]\psi_{lmn} = \left[\partial_t^2-\partial_x^2+V_{l}\right]\psi_{lmn}$ for stationary solutions $\psi_{lmn}$.  After appropriate choice of coordinates it is possible to arrive at similar equations for rotating (Kerr) black holes, where however the potential $\hat{V}$ in the stationary, radial part of the wave equation also becomes frequency dependent~\cite{chandrasekhar_equations_1997,berti_black_2025,PhysRevLett.134.141401}.

The source term $S(x,t)$ relevant for second-order (quadratic) quasinormal modes typically includes, up to derivatives, a product of first-order quasinormal modes. The exact regularized form of the source term can be found elsewhere~\cite{PhysRevD.76.061503}, which also respects the gauge invariant prescription from~\cite{GaugeInvariance}. Also see Refs.~\cite{bucciotti_amplitudes_2024,kerrQNM2}. 
Given all this, our first objective is to cast this general form of wave equation into a Hamiltonian formalism for the temporal modes.  This includes non-linear terms that represent second harmonic generation, which in turn implies squeezing. 

Based on the driven quasinormal mode response predicted by Eq.~\eqref{eqRGS}, we consider a quasinormal mode expansion for part of the second-order solution in the near-field region to make a leading-order estimate for the coupling.  This estimate will be approximate.  Quasinormal modes do not form a complete basis, so the expansion is delicate and non-uniform in space-time.  At a given time shortly past merger it is expected to be most accurate near the source, where most of the amplitude resides, so that the expansion coefficients can be read off (approximately) from behavior in that region.  Careful treatment would require bounding the errors due to spatial cutoff and incompleteness of the basis, and is left for future work.  Our treatment, motivated by the standard approaches to obtaining couplings in the context of normal (zero-damped) modes subject to external forcing, assumes a reasonable cutoff (i.e., neglect of the formally infinite far-field contribution), takes initial mode occupations as input and evolves them for a few cycles to establish equilibrium.  The resulting estimate is far from rigorous, but it supports the semi-phenomenological estimate and indicates how it could be made more precise.

We begin by considering the solution for the homogeneous part of Eq.~\eqref{eqRGS} using the method of separation of variables. The following candidate solution for the homogeneous part of Eq.~\eqref{eqRGS} can be considered, $\psi(x,t)\sim \int d^2\omega \tilde{\psi}(x,\omega)h(t,\omega)$, where $\omega$ is the complex parameter independent of $x,t$ such that $\ddot{h}+\omega^2h =0$ for the temporal part, and,
\begin{equation}
    \left[-\partial_x^2+V(x,\omega)\right]\tilde{\psi}(x,\omega) =\omega^2\tilde{\psi}(x,\omega),\label{qnmF}
\end{equation}
 for the spatial part, which resembles a time-independent Schrodinger equation. Above, we have indicated that the potential $V$ could be frequency dependent, $V=V(x,\omega)$, to relate to the quasinormal modes of Kerr black holes~\cite{chandrasekhar_equations_1997,berti_black_2025,PhysRevLett.134.141401}. Applying the boundary conditions typically pick specific values for the frequencies, $\omega$.  The solutions $\tilde{\psi}(x,\omega)$ in the quasinormal mode context have to satisfy the Siegert boundary conditions~\cite{Siegert}: $\tilde{\psi}(x,\omega)\rightarrow e^{\pm i\omega x}$ as $x\rightarrow \pm\infty$ [in the choice of coordinates, $x\rightarrow -\infty$ corresponds to the event horizon].  
As is well known, such boundary conditions for potentials relevant to the gravitational context imply that the mode frequencies $\omega = \{\omega_n\}$ [for given $l,m$, or generically $\omega = \{\omega_{lmn}\}$] are both discrete and complex, with negative imaginary parts. The corresponding solutions, or quasi-normal modes, are well characterized and understood~\cite{Zerilli,ReggeWheeler,MaggioreQNM,berti_quasinormal_2009,Vishu,Teukolsky,berti_black_2025}, using exact methods or with Wentzel-Kramers-Brillouin (WKB) type approximations~\cite{schutz_black_1985}. Once the solutions are obtained, metric perturbations $h_{\mu\nu}^{(i)},~~i=1,2$ can be reconstructed by following a standardized prescription~\cite{berti_quasinormal_2009,PhysRevLett.134.141401}, while the temporal mode profiles are determined by the discrete quasinormal mode frequencies. 

To estimate the coupling to leading order perturbatively, we consider a particular {\it ansatz} of the form $\psi^{(p)}(x,t)\approx \sum_n \tilde{\psi}_n(x,\omega_n)h_n^{(p)}(t)$ to describe a part of the second-order solution, which can be understood as the driven response of the first order quasinormal modes to the leading order. Such quasinormal mode expansions are generically not complete, and the applicability is best limited to finite regions of spacetime~\cite{Completeness,szpak_quasinormal_2004,settimi_quasi-normal-modes_2009}. By restricting to near-field and quasinormal mode expansions, we mainly miss out on the asymptotic (far-field) behavior of the source and the response. 

The approximations we make here are quite analogous to defining a Fabry-Perot-like cavity of finite length in an optical setting, with open boundary conditions on both ends. Here, with an appropriate definition of an inner product, quasinormal mode expansions are complete within the cavity, and can be used to make meaningful predictions comparable to a normal mode setting~\cite{settimi_quasi-normal-modes_2009}. For readers interested in a complete description of the solution at second order, we recommend that earlier works using Green's functions~\cite{perrone_non-linear_2024,green2,green3,szpak_quasinormal_2004,PhysRevD.107.044040,Leaver,PhysRevD.76.084007,PhysRevD.76.061503,Completeness} are more appropriate.

It is instructive to first consider the limit when the source term reduces to zero to understand the unperturbed sector. Substituting the expansion $\psi(x,t)=\sum_n \tilde{\psi}(x,\omega_n)h_n(t)$, the homogeneous part of the wave equation can be expressed as,
\begin{equation}
    \sum_n \tilde{\psi}_n(x,\omega_n)\left[
\ddot{h}_n(t)+\omega_n^2h_n(t)\right] =0,\
\end{equation}
where we have used Eq.~\eqref{qnmF}. Knowing the solutions for the temporal part of quasinormal modes of the form,
\begin{eqnarray}
    h_n(t)\sim e^{-i\omega_n t},~~\text{where}~~ \omega_n=\pm \omega_{R,n}-i\omega_{I,n},
\end{eqnarray}
we can also write,
\begin{equation}
    \sum_n \tilde{\psi}_n(x,\omega_n)\left[
\ddot{h}_n(t)+2\omega_{I,n}\dot{h}_n(t)+\Omega_n^2h_n(t)\right] =0.\label{modeDc}
\end{equation}
where $\Omega_n^2=\omega_{R,n}^2+\omega_{I,n}^2$ and $\omega_{I,n}\geq 0$ in our notation. Here, focusing on individual terms of the sum, it is evident that the temporal domain of the homogeneous part of the wave equation describes damped quantum harmonic oscillators, with damping coefficients $\gamma_n = 2\omega_{n,I}$ and proper frequencies $\Omega_{n}=\sqrt{\omega_{R,n}^2+\omega_{I,n}^2}$. An insightful discussion of this correspondence can be found in Ref.~\cite{MaggioreQNM}. Here, it is also evident that the individual quasinormal modes in an effective quantum description could be thought of as damped quantum harmonic oscillators. The effect of incorporating the source term can then be understood as forcing. In the temporal domain, the system can be described as a collection of forced, damped, quantum harmonic oscillators, where, given the temporal dependence of the source term as quadratic in first-order modes, each of the modes in the above sum contributes to the second harmonic mode. To second order, each of these forced harmonic oscillators will be vibrating at the frequency of the forcing term. 

\subsection{The driven quasinormal mode response\label{forcing}}

To account for the effect of the source term as small, driving effects on the first-order modes, we decompose the second-order response and the spatial part of the source term $S$, in terms of the first-order mode functions, $\{\tilde{\psi}_n\}$, in the near-field region where we aim to estimate the leading order coupling. While this can be achieved in different problems where $\{\psi_n\}$ form a complete basis within the region of interest, such a decomposition can only account for parts of the solution and source term ``parallel" (in the sense of an appropriate projection) to the quasinormal mode functions since quasinormal modes do not form a complete basis. However, the parts expandable in terms of the mode functions $\{\psi_n\}$ in finite regions of interest is expected to capture the driven quasinormal mode response as discussed below.

The part of the source,  $S_\parallel$, which can be expanded within the region of interest in terms of $\{\psi_n\}$ can be identified from the source $S$ by an appropriate projection method, using a projection operator $\mathcal{L}(\tilde{\psi}_n)[\cdot]$ satisfying $\mathcal{L}(\tilde{\psi}_n)[\tilde{\psi}_m]\propto\delta_{nm}$ for a given region, $\mathcal{C}$. We discuss an exemplary choice for the projection operator in Sec.~\ref {projectionmethod}, valid for any contour $\mathcal{C}$. A norm $\mathcal{N}_n^{2}$ can also be defined using the projection operator $\mathcal{L}(\tilde{\psi}_n)[\cdot]$ as $\mathcal{N}_n^{2}:=\mathcal{L}(\tilde{\psi}_n)[\tilde{\psi}_n]$ (also see Ref.~\cite{PhysRevLett.134.141401}). In what follows, we take $\mathcal{C}$ to be the near-field---vicinity of the peak of the potential barrier. Other alternatives for $\mathcal{L}(\tilde{\psi}_n)[\cdot]$ are also certainly possible, and our prescription here is largely agnostic to the particular choice we make in Sec.~\ref {projectionmethod}.  This is because, using such a projection operator within the region of interest, we can write,
\begin{equation}
    S=S_\parallel+S_\perp,~~\text{where}~~S_\parallel:=\sum_n \mathcal{N}_n^{-2}\mathcal{L}(\tilde{\psi}_n)[S]\tilde{\psi}_n.\label{eqSd}
\end{equation}
 From Eq.~\eqref{eqSd}, it also follows that, $\mathcal{L}(\tilde{\psi}_m)[S_\parallel]=\mathcal{L}(\tilde{\psi}_m)[S]$. Above, $S_\perp = S-S_\parallel$, and it can be noted that $\mathcal{L}(\tilde{\psi}_n)[S_\perp]=0~~\forall n$. It is also evident that $S_\parallel$ is the part whose spatial mode profile can be expanded in terms of the first-order modes, $\tilde{\psi}_n$, within the region $\mathcal{C}$. Naturally, if $\{\psi_n\}$ were to form a complete basis for the problem or the region of interest, $S_\parallel = S$ and $S_\perp = 0$.

In the quasinormal mode context, a complete description, including the tails, must account for the entire source term, and this is studied in Refs.~\cite {PhysRevLett.130.081401,PhysRevLett.130.081402,PhysRevD.76.084007,PhysRevD.76.061503,green3,PhysRevD.107.044040,perrone_non-linear_2024}.  Owing to the lack of completeness for the first order quasinormal modes to describe either the second-order source term or the solution, the complete particular solution must be written as $\phi^{(p)}(x,t)= \psi^{(p)}(x,t)+\chi^{(p)}(x,t)$, where $\chi^{(p)}(x,t)$ satisfies,
\begin{eqnarray}
    \left[\partial_t^2-\partial_x^2+V\right]\chi^{(p)}=S_\perp.\label{eqsp}
\end{eqnarray}  

As the driven quasinormal mode response is sourced by $S_\parallel$ to the leading order, we can reduce the dynamics to that of  driven oscillators from the following wave equation,
\begin{equation}
    \left[
\partial_t^2-\partial_x^2+V\right]\psi^{(p)} \approx S_\parallel.\label{newS}
\end{equation} 
Here, $\psi^{(p)}(x,t)\approx \sum_n \tilde{\psi}_n(x,\omega_n)h_n^{(p)}(t)$ and $S_\parallel$ also permits a mode-expansion in terms of the first-order quasinormal modes, $\tilde{\psi}_n(x,\omega_n)$. Using the above ansatz, we can rewrite Eq.~\eqref{newS} as, 
\begin{equation}
    \sum_n \tilde{\psi}_n(x,\omega_n)\left[
\ddot{h}_{n}^{(p)}(t)+\omega_n^2h_{n}^{(p)}(t)\right] =S_\parallel.\label{modeExp}
\end{equation} 
We have again made use of Eq.~\eqref{qnmF} in Eq.~\eqref{modeExp}, and the latter is an abbreviation for,
\begin{equation}
    \sum_n \tilde{\psi}_{lmn}(x,\omega_{lmn})\left[
\ddot{h}_{lmn}^{(p)}(t)+\omega_{lmn}^2h_{lmn}^{(p)}(t)\right] =S_{\parallel,lm},\label{modeExp2}
\end{equation} after silencing $l,m$ indices. For our considerations, we restrict to the $l=4,m=4$ leading second-order quasinormal modes sourced by the $l=2,m=2$ mode. This allows us to project out the spatial part on both ends and reduce the equations to that of a forced damped harmonic oscillator in the temporal domain, from which an interaction Hamiltonian can be derived. 

The source term in the near-field region to leading order is approximated as $S\approx S_{\parallel}$ -- that is, a superposition of the spatial part of the first order modes,  $\tilde{\psi}_n(x,\omega)$. Physically, this provides an approximate description of the spatial part of the second harmonic in the near-field, as it is produced by forcing the first-order quasinormal modes to the second harmonic. 

\subsection{Projection to quasinormal modes\label{projectionmethod}}
To pick out the response of individual quasinormal modes, we need to consider some type of orthogonality relations for quasinormal modes. This is not straightforward, because the spatial modes of quasinormal modes are not square-integrable. Several alternatives  that respect the ``quasinormal" aspect of the modes have been considered in the literature~\cite{comp1,comp2,comp3,comp4,comp5,comp6,PhysRevLett.134.141401,sauvan_normalization_2022,Leaver,kristensen_modeling_2020}. They explore different approaches including include the use of Green's functions~\cite{Leaver,comp1} and bi-orthogonal bases. Here one finds states $\{|\phi_j\rangle\}$ that are orthogonal to $\{|\tilde{\psi}_n\rangle\}$ such that the inner product $\langle \phi_j|\tilde{\psi}_n\rangle\propto\delta_{jn}$. Electromagnetic/optical analogies are discussed in Refs.~\cite{kristensen_modeling_2020,leakyOptic,settimi_quasi-normal-modes_2009}.

Here we discuss an exemplary choice for the projection operator, which is motivated by treatment in Refs.~\cite {leakyOptic,comp1,PhysRevLett.134.141401,comp2}. This by no means is expected to be a unique choice. Alternatives are certainly possible, and our analysis only assumes the availability of such a projection operator for quasinormal modes. To define the projection operator, we begin with the abbreviated notation $\psi_{j}=\tilde{\psi}(\omega_j,x).$ It follows from Eq.~\eqref{qnmF} that,
\begin{equation}
    \psi_j\left[-\frac{\partial}{dx^2}+V(x,\omega_n)\right]\psi_n = \omega_n^2\psi_j\psi_n~~\text{and}~~\psi_n\left[-\frac{\partial}{dx^2}+V(x,\omega_j)\right]\psi_j = \omega_j^2\psi_j\psi_n,
\end{equation}
where we have considered the generalized scenario where $V$ is $\omega$ dependent. Taking the difference of the two equations for $n\neq j$, we obtain,
\begin{equation}
    \psi_j\frac{\partial}{dx^2}\psi_n-\psi_n\frac{\partial}{dx^2}\psi_j = \left[U(x,\omega_j)-U(x,\omega_n)\right]\psi_j\psi_n,~~U(x,\omega_k) = \omega_k^2-V(x,\omega_k).
\end{equation}
Multiplying with $(\omega_j^2-\omega_n^2)^{-1}$ on both sides and and integrating over a contour $\mathcal{C}$,
\begin{eqnarray}
    &&\int_{\mathcal{C}} dx \frac{\left[U(x,\omega_j)-U(x,\omega_n)\right]}{(\omega_j^2-\omega_n^2)}\psi_j\psi_n\nonumber\\&=&\frac{1}{(\omega_j^2-\omega_n^2)} \left\{\bigg[\psi_j\frac{\partial}{dx}\psi_n-\psi_n\frac{\partial}{dx}\psi_j \bigg]\bigg|_{\partial \mathcal{C}}- \int_{\mathcal{C}} dx \frac{\partial}{dx}\psi_j\frac{\partial}{dx}\psi_n+\int_{\mathcal{C}} dx \frac{\partial}{dx}\psi_j\frac{\partial}{dx}\psi_n\right\}. 
\end{eqnarray}
The last two terms cancel, and we have (for $n\neq j$),
\begin{eqnarray}
    \int_{\mathcal{C}} dx~ \frac{\left[U(x,\omega_j)-U(x,\omega_n)\right]}{(\omega_j^2-\omega_n^2)}\psi_k\psi_n -\frac{1}{(\omega_j^2-\omega_n^2)}\bigg[\psi_j\frac{\partial}{dx}\psi_n-\psi_n\frac{\partial}{dx}\psi_j  \bigg]\bigg|_{\partial \mathcal{C}}= 0.
\end{eqnarray}
More formally, we can write as,
\begin{eqnarray}
    \lim_{\omega_k\rightarrow\omega_j}\left\{\int_{\mathcal{C}} dx~ \frac{\left[U(x,\omega_k)-U(x,\omega_n)\right]}{(\omega_k^2-\omega_n^2)}\psi_k\psi_n -\frac{1}{(\omega_k^2-\omega_n^2)}\bigg[\psi_k\frac{\partial}{dx}\psi_n-\psi_n\frac{\partial}{dx}\psi_k  \bigg]\bigg|_{\partial \mathcal{C}}\right\}= 0,\label{innnm}
\end{eqnarray}
when $n\neq j$. As a formal device, the limit $\omega_k\rightarrow \omega_j$ can be taken continuously along an appropriate contour in the complex plane of quasinormal mode frequencies, exploiting the circumstance that the frequencies $\{\omega_n\}$ (as well as $U(x,\omega_k)$ and $\psi_k$), although discrete for a given black hole, depend on continuously variable black hole parameters such as the black hole mass and spin. Hence, the intermediate values of $\omega_k$ are also physically meaningful (also see Ref.~\cite{PhysRevLett.134.141401}). With the identification, $H=[-\partial_x^2+V(x,\omega)]$, we see from Eq.~\eqref{qnmF} that $H\psi_j=\omega_j^2\psi_j.$ This allows us to define the following operator (where the operations are performed in the order they appear),
\begin{eqnarray}
    L(\psi_j)[\cdot]&:=&\lim_{\omega_k\rightarrow\omega_j}\left\{\int_{\mathcal{C}} dx~ \psi_k \left[U(x,\omega_k)+\partial_x^2\right]\frac{1}{(\omega_k^2-H)}[\cdot] -\bigg[\psi_k\frac{\partial}{dx}-\bigg(\frac{\partial}{dx}\psi_k\bigg)  \bigg]\frac{1}{(\omega_k^2-H)}[\cdot]\bigg|_{\partial \mathcal{C}}\right\},\nonumber\\
\end{eqnarray}
which obeys $L(\psi_j)[\psi_n] = 0$ when $j\neq n$ [follows from Eq.~\eqref{innnm}]. Hence, the operator is very convenient to project out quasinormal modes other than $\psi_j$ from a linear expansion. Proceeding with the Schwarzschild scenario, we also note that
\begin{eqnarray}
    L(\psi_j)[\psi_j]&=&\lim_{\omega_k\rightarrow\omega_j}\int_{\mathcal{C}} dx~ \frac{\left[U(x,\omega_k)-U(x,\omega_m)\right]}{(\omega_k^2-\omega_j^2)}\psi_k\psi_j \nonumber\\&-&\lim_{\omega_k\rightarrow\omega_j}\bigg[\psi_k\frac{\partial}{dx}\psi_j-\bigg(\frac{\partial}{dx}\psi_k\bigg)\psi_j  \bigg]\frac{1}{(\omega_k^2-\omega_j^2)}\bigg|_{\partial \mathcal{C}}.
\end{eqnarray}
The first term reduces to,
\begin{eqnarray}
    \lim_{\omega_k\rightarrow\omega_j}\int_{\mathcal{C}} dx~ \frac{\left[U(x,\omega_k)-U(x,\omega_j)\right]}{(\omega_k^2-\omega_j^2)}\psi_k\psi_j = \int_{\mathcal{C}} dx~ \frac{dU(x,\omega_k)}{d\omega_k^2}\Bigg|_{\omega_k=\omega_j}\psi_j^2.
\end{eqnarray}
To evaluate the second term, we may define [recalling the abbreviated notation $\psi_{j}=\tilde{\psi}(\omega_j,x)$ for clarity],
\begin{eqnarray}
    g(\omega_k) = \tilde{\psi} (\omega_k,x)\frac{\partial}{\partial x}\tilde{\psi}(\omega_j,x) -\bigg[\frac{\partial}{\partial x}\tilde{\psi}(\omega_k,x)\bigg]\tilde{\psi}(\omega_j,x). 
\end{eqnarray}
Note that, $g(\omega_j) = 0$. This yields,
\begin{eqnarray}
    \lim_{\omega_{k}\rightarrow\omega_j}\frac{g(\omega_k)-g(\omega_j)}{\omega_k^2-\omega_j^2}=\lim_{\omega_{k}\rightarrow\omega_j}\frac{g(\omega_k)}{\omega_k^2-\omega_j^2}=\frac{\partial}{d\omega_{k}^2}g(\omega_k)\bigg|_{\omega_k=\omega_j}.
\end{eqnarray}
Hence we obtain,
\begin{eqnarray}
    L(\psi_j)[\psi_j]=\int_{\mathcal{C}} dx~ \frac{dU(x,\omega_k)}{d\omega_k^2}\Bigg|_{\omega_k=\omega_j}\psi_j^2 -\frac{\partial}{d\omega_k^2}\bigg[\psi_k\frac{\partial}{dx}\psi_j-\bigg(\frac{\partial}{dx}\psi_k\bigg)\psi_j  \bigg]_{\partial \mathcal{C}}\Bigg|_{\omega_k=\omega_j} := \mathcal{N}_j^2,
\end{eqnarray}
where the norm $\mathcal{N}_{j}^2$ is complex in general~\cite{PhysRevLett.134.141401,leakyOptic,comp1}. 
\subsection{Estimating the driving force\label{forceqn}}
The action of an appropriate projection operator,  $\mathcal{L}(\psi_j)[\cdot]$, on the left-hand side of Eq.~\eqref{modeExp} only retains the coefficient of the $j^{\text{th}}$ mode in the expansion, with the normalization factor $\mathcal{N}_{j}^2$. The action of $\mathcal{L}(\psi_j)[\cdot]$ on the right-hand side of Eq.~\eqref{modeExp} must therefore determine the driving term of the wave equation with appropriate pre-factors. Hence, we can write,
\begin{eqnarray}
 &&\left[\ddot{h}_{j}^{(p)}(t)+\omega_{j}^2h_{j}^{(p)}(t)\right] =\mathcal{N}_{j}^{-2}\mathcal{L}(\psi_j)[S_\parallel] =\mathcal{N}_{j}^{-2}\mathcal{L}(\psi_j)[S] =g_{j}(t),\label{sourD}
\end{eqnarray}
where we have also used Eq.~\eqref{eqSd}.
In the gravitational context, we know that the temporal part of the source contains the product of lower-order quasinormal modes, dominated by the fundamental ($\omega_{220}\equiv \omega$) mode~\cite{PhysRevD.76.061503}. More generically, we can approximate, to the leading order, $g_j(t) \approx \sqrt{2}\Lambda_j [h_{\omega}(t)]^2$, where we have denoted $h_{\omega}$ as the fundamental mode having exactly half the frequency to that of the second harmonic mode $h_{j}^{(p)}(t)$. Recall that here the superscript $(p)$ denotes that $h_{j}^{(p)}(t)$ is the particular solution to Eq.~\eqref{sourD}. We have also absorbed any pre-factors to the effective parameter $\Lambda_j$, determined by Eq.~\eqref{sourD}. While these results apply to any of the modes, we restrict ourselves to the long-lived modes that have a substantial real frequency (propagating) part, which are also the ones likely to be observed. We thus obtain,
\begin{equation}
\ddot{h}_{j}^{(p)}(t)+\omega_{j}^2h_{j}^{(p)}(t) \approx \sqrt{2}\Lambda_j [h_{\omega}(t)]^2,\label{eqTemp}
\end{equation}
where $h_{\omega}$ satisfies $\ddot{h}_{\omega}+\omega^2h_{\omega}=0$.  
Multiplying by an effective moment of inertia $\mu_j\sim c^5/(G\omega_j^3)$ across, the equation resembles the equation for a forced harmonic oscillator, except note that $\{\omega_j\}$ and potentially $\Lambda_j$, are complex. By considering long-lived modes, to the leading order, we may treat these parameters to be real and neglect the contributions from their imaginary parts. This only applies to lower-order (small $j$) quasinormal modes, as the higher-order ones are increasingly damped; however, the lower-order modes are also the ones of interest here.

Re-instating the notations in the Schwarzchild context as an example, to the leading order, we have,
\begin{equation}
\ddot{h}_{44j}^{(p)}(t)+\omega_{44j}^2h_{44j}^{(p)}(t) \approx \sqrt{2}\Lambda_j [h_{220}(t)]^2~~\text{and}~~\ddot{h}_{kkj}(t)+\omega_{kkj}^2h_{kkj}(t)=0,~~k\in\{2,4\}.\label{eqpart}
\end{equation}
With $k=4$ in the second equation, it is, of course, the homogeneous part of the first equation. 
This motivates the suggestive form for the coupled wave equation Eq.~\eqref{eqSeco} used in the main text.  

\section{Exact analysis of squeezing in second harmonic generation\label{AppSHG}}

Here we briefly revisit a perturbatively exact treatment to estimate the degree of squeezing in second harmonic generation by closely following the analysis in Refs.~\cite {OuSHG,PhysRevA.49.2157,suhara_theoretical_1996}, which the readers may refer to for details. In comparison, our analysis is in the temporal domain. Consider the following dynamical equations that describe second harmonic generation,
\begin{equation}
    \frac{dA_{\omega}}{dt} =-\lambda  A^\dagger_\omega A_{2\omega},~~\frac{dA_{2\omega}}{dt} = \frac{\lambda}{2}A_{\omega}^2.
\end{equation}
Following the standard practice, one can define the quantum operators about the classical mean-field solutions,
\begin{equation}
    a_{i\omega} = A_{i\omega}-\langle A_{i\omega}\rangle,~~i=1,2. 
\end{equation}
They obey,
\begin{equation}
    \frac{d\langle A_{\omega}\rangle}{dt} =-\lambda  \langle A_\omega\rangle^* \langle A_{2\omega}\rangle,~~\frac{d\langle A_{2\omega}\rangle }{dt} = \lambda \langle A_{\omega}\rangle^2.
\end{equation}
and
\begin{equation}
    \frac{da_{\omega}}{dt} =-\lambda  (a^\dagger_\omega \langle A_{2\omega}\rangle +a_{2\omega} \langle A_{1\omega}\rangle^*),~~\frac{da_{2\omega}}{dt} = \lambda \langle A_{\omega}\rangle a_\omega.
\end{equation}
One can now define the following change of variables~\cite{PhysRevA.49.2157,OuSHG},
\begin{eqnarray}
    u_{\omega}=\frac{|\langle A_{\omega}(t)\rangle|}{|\langle A_{\omega}(0)\rangle|},~~u_{2\omega}=\frac{\sqrt{2}|\langle A_{2\omega}(t)\rangle|}{|\langle A_{\omega}(0)\rangle|},~~\text{and}~~\tau = \frac{1}{\sqrt{2}}\lambda |\langle A_{\omega}(0)\rangle|t,
\end{eqnarray}
in terms of which the mean-field equations are recast as,
\begin{equation}
    \frac{du_{\omega}}{d\tau} = -u_{\omega}u_{2\omega},~~    \frac{du_{2\omega}}{d\tau} = u_{\omega}^2,
\end{equation}
where perfect phase-matching is assumed, $\text{Arg}(\langle A_{2\omega}(t)\rangle) = \phi_{2\omega}(t) = 2\text{Arg}(\langle A_{\omega}(t)\rangle)=2\phi_{\omega}(t).$ Hence, only the real amplitudes come into play. The system of equations admit solutions (assuming initial conditions $u_{\omega}(0) = 1$, and $u_{2\omega}(0) = 0$)~\cite{PhysRevA.49.2157,OuSHG},
\begin{equation}
    u_{\omega}(\tau)=\text{sech}\tau,~~u_{2\omega}(\tau)=\tanh\tau.
\end{equation}
For the generalized position and momentum quadratures, 
$\hat{X}_{j\omega}(\tau)=\frac{1}{\sqrt{2}}(a_{j\omega}e^{-i\phi_{j\omega}}+a^\dagger_{j\omega}e^{i\phi_{j\omega}})$ and $\hat{P}_{j\omega}(\tau)=\frac{1}{i\sqrt{2}}(a_{j\omega}e^{-i\phi_{j\omega}}-a^\dagger_{j\omega}e^{i\phi_{j\omega}}),$
one obtains equations~\cite{PhysRevA.49.2157,OuSHG},
\begin{eqnarray}
    \frac{d\hat{X}_{\omega}}{d\tau} &=& -\hat{X}_{\omega}\tanh\tau-\sqrt{2}\hat{X}_{2\omega}\text{sech}\tau,\nonumber\\\frac{d\hat{X}_{2\omega}}{d\tau}&=&\sqrt{2}\hat{X}_{\omega}\text{sech}\tau\nonumber\\
     \frac{d\hat{P}_{\omega}}{d\tau} &=& \hat{P}_{\omega}\tanh\tau-\sqrt{2}\hat{X}_{2\omega}\text{sech}\tau,\nonumber\\\frac{d\hat{P}_{2\omega}}{d\tau}&=&\sqrt{2}\hat{P}_{\omega}\text{sech}\tau .
\end{eqnarray}
Solutions are straightforward. One obtains~\cite{PhysRevA.49.2157,OuSHG},
\begin{eqnarray}
    \hat{X}_{\omega}(\tau)&=&(1-\tau \tanh\tau)\hat{X}_{\omega}(0)\text{sech}\tau-\sqrt{2}\hat{X}_{2\omega}(0)\tanh\tau\text{sech}\tau,\nonumber\\\hat{X}_{2\omega}(\tau)&=&\frac{1}{\sqrt{2}}(\tanh\tau +\tau \text{sech}^2\tau)\hat{X}_{\omega}(0)+\hat{X}_{2\omega}(0)\text{sech}^2\tau,\nonumber\\
    \hat{P}_{\omega}(\tau)&=&\hat{P}_{\omega}(0)\text{sech}\tau-\frac{1}{\sqrt{2}}(\sinh\tau +\tau \text{sech}\tau)\hat{P}_{2\omega}(0),\nonumber\\
    \hat{P}_{2\omega}(\tau)&=&\sqrt{2}\hat{P}_{\omega}(0)\tanh\tau+(1-\tau\tanh\tau)\hat{P}_{2\omega}(0).
\end{eqnarray}
Based on this, one finds the covariance matrix elements (for initial conditions where $\langle \hat{X}^2_{j\omega}(0)\rangle=\langle \hat{P}^2_{j\omega}(0)\rangle=\frac{1}{2}$ and cross terms initially zero),
\begin{eqnarray}
    \langle \hat{X}^2_{j\omega}(\tau)\rangle =  \langle \hat{X}^2_{j\omega}(0)\rangle S_{j\omega X},~~\text{and}~~\langle \hat{P}^2_{j\omega}(\tau)\rangle =  \langle \hat{P}^2_{j\omega}(0)\rangle S_{j\omega P},
\end{eqnarray}
where~\cite{PhysRevA.49.2157,OuSHG}:
\begin{eqnarray}
    S_{\omega X} &=&(1-\tau\tanh\tau)^2\text{sech}^2\tau+2\tanh^2\tau \text{sech}^2\tau,\nonumber\\
      S_{2\omega X} &=&\frac{1}{2}(\tanh\tau+\tau\text{sech}^2\tau)^2+\text{sech}^4\tau,\nonumber\\
    S_{\omega P} &=&\frac{1}{2}(\sinh\tau+\tau\text{sech}\tau)^2+\text{sech}^2\tau,\nonumber\\  
     S_{2\omega P}&=&2\tanh^2\tau+(1-\tau\tanh\tau)^2.
\end{eqnarray}
For large $\tau$, note that, $S_{\omega X}\rightarrow 4\tau^2 e^{-2\tau}$, $ S_{2\omega X}\rightarrow 1/2$, $S_{\omega P}\rightarrow e^{2\tau}/8$, and $S_{2\omega P}\rightarrow \tau^2$~\cite {OuSHG,PhysRevA.49.2157,suhara_theoretical_1996}, suggesting the exponential enhancement in the noise of the phase quadrature. While the predictions are limited by the validity of the linearized equations that, for instance, assume a highly intense coherent field on the fundamental mode to start with, we can take these results to indicate that substantial squeezing can, in principle, be generated via the second harmonic generation in the fundamental mode. However, for the scenarios we are interested in, we have also excluded the decay, or damping, of the quasinormal modes, so our predictions are relevant for weakly damped quasinormal modes that are long-lived, applicable at timescales much smaller than the lifetime of quasinormal modes. Hence, the relevant degree of squeezing can be estimated based on the limit of small $\tau$, captured already by the perturbative expansion, Eq.~\eqref{pertSqueeze} of the main text.   

\bibliography{refer}
\end{document}